\documentstyle[preprint,aps]{revtex}

\tighten
\begin{document}

\draft

\title{
         Structure of the inner crust of neutron stars: crystal lattice or
         disordered phase ?}

\author     {
             P. Magierski$^{1}$ and P.-H. Heenen$^{2}$ \\
         $^1${\em Institute of Physics, Warsaw University of Technology,}\\
             {\em ul. Koszykowa 75, PL--00662, Warsaw, Poland} \\
         $^2${\em Service de Physique Nucl\'{e}aire Th\'{e}orique,} \\
             {\em U.L.B - C.P. 229, B 1050 Brussels, Belgium} \\
            }

\maketitle

\begin{abstract}
We investigate the inner crust structure of neutron stars using
the Skyrme-Hartree-Fock approach with the Coulomb interaction
treated beyond the Wigner-Seitz approximation.
Our results suggest that the shell effects associated with unbound neutrons
play an important role and, in particular,
lead to complicated phase transition pattern
between various nuclear phases (as a function of the density).
Namely, we show that the relative energies of different phases
are rapidly oscillating functions of the neutron density.
In the semiclassical approach
this behavior is explained as an interference effect due
to periodic orbits of similar lengths.
We discuss also the dependence of the shell effects on pairing
correlations.
\end{abstract}

\pacs{PACS numbers: 97.60.Jd, 26.60.+c, 21.65.+f, 21.60.Jz}
\narrowtext

\section{INTRODUCTION}

The outer layer of a neutron star has a complex  structure which depends
strongly on the nuclear density.
In the inner crust of the star,
due to the high density and pressure, a large fraction of neutrons
occupies unbound states. Nuclei which are still present are therefore
immersed in a neutron gas.
The structure of this part of the star has been the subject
of a considerable theoretical effort
\cite{bbp,bba,rbp,arp,nvb,rpw,hsy,ohy,lpr,wko,lfb,oya1,lor,oya2,pra,dhm,dha}.
The main conclusion emerging from these studies is
that the nuclei  form exotic phases, stabilized
by the Coulomb interaction, and
characterized by various crystal lattice structures.

The theoretical predictions on the structure and stability of the nuclear
phases are based on the  minimization of the energy
of the neutron-proton-electron (npe) matter within a Wigner-Seitz cell,
whose radius is treated as a variational parameter.
An agreement has been
reached on the existence of five phases formed in the region where
the nucleon density varies from $0.03$ to $0.1$ fm$^{-3}$.
The following chain of phase transitions is predicted
as the density increases: spherical nuclei $\rightarrow$ rods
$\rightarrow$ slabs $\rightarrow$ tubes $\rightarrow$ bubbles $\rightarrow$
uniform matter. The physical mechanism responsible for
the phase transition pattern results from the interplay
between the Coulomb and surface energies.
However, the differences between the energy density of the various
phases are only of the order of a few $keV/fm^{3}$, for a wide range of
densities \cite{pra}.
The precise transition pattern is therefore dependent on the
ingredients of the theoretical models. For example, recent calculations
based on the self-consistent Hartree-Fock (HF) method
have predicted that the phase composed of spherical nuclei
is stable up to a density around $0.077$ fm$^{-3}$, above which
the uniform matter is energetically favored \cite{dha}.
Other phases  appear at higher energies and form excited configurations.

In the present paper we focus on an effect which has been
overlooked in previous studies but should strongly influence the
phase transition pattern. In all to date approaches,
shell effects have been either completely
neglected, as in  models based on a liquid drop formula or in
Thomas-Fermi models, or have been taken into account only for bound nucleons.
On the other hand, it has been recently shown in
the semiclassical approximation, based on the Gutzwiller trace formula,
that the shell effects in  inhomogeneous nuclear matter are quite strong and
yield corrections to the energy density
of the order of several $keV/fm^{3}$. This value is
comparable to the differences between the liquid drop energy densities
predicted for the different nuclear phases \cite{bma1,bma2,bma3}.
Moreover, the shell energy oscillates as a
function of  density and depends strongly on the
spatial order in the system. In particular, its amplitude
depends both on the nuclear geometry and on the lattice type.

The shell energy of npe matter in the
neutron star crust has two origins. The first one is associated
with bound nucleons and has been shown to be small
by Oyamatsu and Yamada in Ref. \cite{oya2}.
The second one is related to unbound neutrons which may scatter on
inhomogeneities and form resonant states. The situation is somewhat
similar as for the Casimir effect in quantum field theory and condensed
matter (see e.g. Refs. \cite{cas,fish,kgo} and references therein),
where, in some systems, a fluctuation induced interaction leads to a
correction to the total energy.
In the same way, the scattering of unbound neutrons
lead to an effective interaction between nuclei
immersed in the neutron environment. This effect
can be termed the fermionic Casimir effect and has been studied
for simple geometries in Refs. \cite{bma1,bma2,bma3,bma4,bwi}.
For impurities that are small compared
to the neutron Fermi wavelength, the generated interaction
is mainly due to s-wave scattering. It gives then rise to
the Ruderman-Kittel correction which is small compared to the
Fermi energy of neutrons \cite{rki,fwa}.

However, the s-wave scattering limit is not valid in the neutron star crust,
since the neutron Fermi momentum $k^{n}_{F}$ is of
the order of $1 fm^{-1}$ and thus $k^{n}_{F}R > 1$, where $R$ is the size of
an inhomogeneity. Consequently the contribution of higher partial waves
to the scattering process cannot be neglected, and gives rise
to much larger energy correction, as for large objects more of the incident
wave will be reflected \cite{bma1,bma2,bma3,bma4,bwi}.
The leading shell energy contribution,
associated with the neutron scattering, to
the total energy can be determined in the semiclassical approximation
for various geometries of nuclear
shapes (see  appendix B).
It has been shown to yield  energy
corrections that influence the structure of inhomogeneous nuclear matter
\cite{bma1,bma2,bma3}. It is also obvious that since the gross shell structure
is determined by the shortest periodic orbits in the system, the shell energy
is a sensitive function of the crystal lattice type, as well as
of the shape of
inhomogeneities \cite{bma1,bwi}.  Unfortunately the semiclassical approach has
drawbacks. First, it assumes that the obstacles are impenetrable
scatterers which may overestimate the amplitude of the shell energy.
Second, the method lacks of the mutual interplay between the shell energy
and the liquid drop energy. Such a coupling is quite important since it
allows a part of the shell energy to be ``absorbed'' into deformation.
Hence a microscopic treatment of the problem is needed where both effects,
i.e. the one coming from the liquid drop energy
and the shell energy term are treated on the same footing.

\section{THE METHOD}

The ground  and excited (isomeric) states of npe matter
have been determined by the Skyrme-Hartree-Fock method. We have solved
the HF equation by discretization in coordinate space,
within a rectangular box and with  periodic
boundary conditions imposed on the nucleon wave functions \cite{bfh}.
This technique has the advantage that it allows to describe any kind
of shapes of the nuclear density distribution with the same accuracy, and to
introduce naturally  periodic boundary conditions. The code
that was previously developed to study nuclei with an ellipsoidal
symmetry has been modified to accommodate the periodic solutions.

The high efficiency of the method can be attributed to the
iterative procedure used to solve the HF equations, the imaginary time step
method \cite{dfk}. One of its features is that only
those single-particle wave functions which contribute to the
mean field are calculated. It is particularly important
in our case, since the densities which we are dealing with here require to
consider up to $1000$ nucleons per box. In order to take them
into account properly, we have used $600$ wave functions.
Due to Kramers degeneracy, they correspond to $1200$ single-particle states
and thus form a sufficiently large Hilbert space for our purposes.

As a starting point of the HF iterations, Fermi gas wave functions have
been used, which turned out to provide a good first guess of the true
ground state. The criterion for the termination of the iteration
process is the total energy difference obtained after a fixed
number of imaginary time steps.  In our calculations
we terminated the iteration procedure when the
differences between the total energy calculated
at two successive iterations did not exceed $1 keV$. For stable
configurations this value was of an order of magnitude lower.

An important parameter of a mesh is the spacing between the mesh points.
It must be small enough to lead to accurate results and large enough
to accommodate the very large number of wave functions needed to describe
the neutron gas. We have chosen as a compromise between these two
requirements a value of the mesh size equal to $\Delta x=1.3 fm$.

This choice of $\Delta x$ fixes a bound to
the maximum momentum of a nucleon that can be represented by the
discretized wave function. The highest  momentum is indeed equal to
$k_{crit}=\pi/\Delta x\approx 2.42 fm^{-1} $, and corresponds to
the maximum kinetic energy
$\epsilon_{crit}=
\displaystyle{\frac{\hbar^{2}\pi^2}{2m\Delta x^{2}}}\approx 121 MeV$.

There are two main sources of errors caused
by the discretization. One of them is related to unbound neutrons which
may occupy high energy states and thus can not be properly described
by a discretized wave function.
However even at the highest neutron density that we
are interested in here, i.e. $\rho_{n}\approx 0.1$ fm$^{-3}$,
the neutron Fermi momentum
is equal to $k^{n}_{F}\approx 1.44$ fm$^{-1}$ which is far below the critical
value. A second  source of errors is associated
with nucleons that are bound in a nucleus.
Typically, a nuclear potential at densities $\rho > 0.03$ fm$^{-3}$
has a depth which does not exceed  $20$ MeV and  a radius lower than $10$ fm
\cite{oya1}. In a Fermi gas approximation for bound nucleons,
the momentum of the last bound nucleon is around $1$ fm$^{-1}$,
well below the maximal value that can be represented on the mesh.
Let us note that the Fermi gas approximation gives an upper bound
of the maximum momentum since the wave function
of a bound state has a long tail due to the large diffuseness of a
nuclear potential immersed in a neutron gas.

To decrease the numerical task, we have
assumed that the nuclear density distribution is symmetrical with
respect to three planes, allowing however the existence of triaxial
shapes. Parity breaking configurations are thus excluded.

Different phases may coexist for a given nuclear density. To
determine them and to study their stability requires the
construction of adiabatic paths between phases.
This has been done by introducing a constraint on the
quadrupole moment of the proton density distribution in the Skyrme
HF equations.
In order to compare energies of configurations differing by the
proton quadrupole moment,
a careful treatment of the Coulomb interaction, both
between protons in different nuclei of the lattice, as well
as between protons and electrons, is needed. It has lead us
to solve the Poisson equation for the electric potential
within a box with periodic boundary
conditions. This approach is clearly free from limitations
of the usual Wigner-Seitz approximation, where the influence of the
lattice type on the Coulomb energy is not taken into account.
The electrons are assumed to
form a uniform relativistic gas where the screening effects are neglected.
The total charge of the box is zero\cite{lfb}, which gives a relation
between the electron and proton densities.

Summarizing, we solve the Hartree-Fock equations for the npe
matter with contributions in the energy functional coming from the nuclei,
the neutron gas and the electrons. In such a calculation, the
liquid drop  and the shell energy parts of the energy are automatically
and self-consistently included.
The total energy of the system can be expressed in the form:
\begin{equation} \label{energy}
E_{tot}(\rho_{p},\rho_{n},\rho_{e})=
E_{Skyrme}(\rho_{p},\rho_{n}) +
E_{Coul}(\rho_{p},\rho_{e}) +
E_{el} (\rho_{e}),
\end{equation}
where $E_{Skyrme}$ comes from the Skyrme energy functional \cite{bfh},
$E_{Coul}$ is the Coulomb energy term, and $E_{el}$ denotes
the electron kinetic energy term. The densities
$\rho_{n}, \rho_{p}, \rho_{e}$ denote the average densities
of neutrons, protons and electrons, respectively. They are related
to the density distribution through the equation:
\begin{equation}
\rho_{i}=\frac{1}{V}\int_{V}\rho_{i}({\bf r})d^{3}r ,
\mbox{\hspace{1cm} i=n,p,e},
\end{equation}
where $V$ is the  volume of the cell.
The charge neutrality  condition implies that $\rho_{e}=\rho_{p}$.
The total nuclear density is given by $\rho=\rho_{n}+\rho_{p}$.
In the calculations we have used the SLy4 Skyrme
force. This recent parametrization \cite{cmb}
gives a good description of neutron matter at
low densities \cite{wff,dha}.
For a few test cases, calculations have also been
performed with the SLy7 force which does not lead to any qualitative
differences.

The shell energy which is
implicitly included in the expression (\ref{energy})
can be decomposed into two parts. The first one, denoted by
$E_{shell}^{in}$, is determined by the density of neutrons $\rho_{n}^{in}$
and protons $\rho_{p}^{in}$ inside the nucleus i.e.:
\begin{equation}
\rho_{i}^{in}=\frac{1}{V_{nucl}}\int_{V_{nucl}}\rho_{i}({\bf r})d^{3}r,
\mbox{ \hspace{1cm} i=n,p},
\end{equation}
where $V_{nucl}$ is the volume of the nucleus. The second,
$E_{shell}^{out}$,  is a
function of unbound neutrons $\rho_{n}^{out}$ and
protons $\rho_{p}^{out}$, where
\begin{equation}
\rho_{i}^{out}=\frac{1}{V-V_{nucl}}\int_{V-V_{nucl}}\rho_{i}({\bf r})d^{3}r,
\mbox{ \hspace{1cm} i=n,p}.
\end{equation}
Since $\rho_{p}^{out}$ is practically zero for
$\rho < 0.08$ fm$^{-3}$ \cite{dha}, the
contribution to the energy due to unbound protons is marginal.

It should be emphasized that the decomposition of the density
into an inner and an outer part is somewhat arbitrary.
There are various methods of defining these quantities that have been
used in liquid drop based approaches \cite{pra,dhm}. In the HF method
this decomposition is not necessary to calculate the total energy of
the system. Nevertheless,
in order to understand its behavior as a function of
physical parameters such as nuclear density, proton fraction,
lattice constant, etc., it is important to identify and analyze
the origin of the various contributions to the total energy.

\section{RESULTS}

The HF equations are solved for a fixed
total density $\rho$, box size $d$ and with a constraint
imposed on the proton quadrupole moment.
In this way, it is possible to determine not only the ground
state but also excited configurations,
differing from the ground state by, for example, the lattice type.
Since we solve the problem in a cubic box with three symmetry planes,
several lattice geometries can be generated like e.g. simple
cubic crystal (scc), face centered crystal (fcc), or body centered
crystal (bcc) \cite{pbh}.

In order to verify that the absolute minimum of the functional
(\ref{energy}) has been found,
we have performed a set of constrained calculations corresponding to
various values of the proton quadrupole moment for
each value of the
total nuclear density. It should be emphasized that we did not perform
the total energy minimization with respect to the parameters of the box
which were kept fixed at values corresponding to those obtained in the
reference \cite{oya1}.

In the figure 1 we show the ground state
energy density as a function of the
proton fraction: $Z/A=\rho_p /\rho$
($Z = V\rho_p$ and
$A= V\rho $, where $V=d^3$ is the volume of the box)
for four values of the nucleon density.
The calculations indicate that
for the density range we
consider here, the optimal $Z/A$ ratio varies between $0.03$ and $0.04$.

For a total density
$\rho=0.055$ fm$^{-3}$, the energy density is plotted
in Fig. 2 (middle part of the left panel)
as a function of the proton quadrupole moment $Q_p = Q^{p}_{20}$.
Most studies predict that, at this density,
the nuclear spherical phase is energetically favored
(see e.g. \cite{pra,oya1} and references therein).
One can see  that in our self-consistent calculations,
the energy minimum appears for a non-zero
quadrupole moment. In order to identify the shape of the nuclei,
as well as the lattice type,
we have plotted the neutron and proton density
distributions integrated along the $z$ and $y$ axes:
\begin{eqnarray}
\rho_{i}^{z}(x,y)&=&\int_{0}^{d} \rho_{i}(x,y,z)dz, \\
\rho_{i}^{y}(x,z)&=&\int_{0}^{d} \rho_{i}(x,y,z)dy, \\
\end{eqnarray}
where $i=n,p$. In the lower and upper parts of
the left panel of Fig. 2, the integrated density distributions
are plotted for both protons and neutrons.
We have shown the density distributions for the configurations
indicated by arrows, which correspond to the spherical and the
most elongated nuclear shapes. The density distributions in the
energy minimum (not plotted) have a pattern very similar to the
spherical one. They correspond to a scc lattice
of deformed nuclei immersed in a neutron gas.

An excited minimum appears for
a proton quadrupole moment of about $2500 fm^{2}$. It is associated with
a rod-like (``spaghetti'') phase (upper part of the left panel of Fig. 2).
The  barrier between these minima is of the order of
$0.5 keV/fm^{3}$. One can see that contrary to  previous
calculations, the spherical phase is no longer the most favored
one and appears at an excitation energy of about $1 keV/fm^{3}$, even above
the ``spaghetti'' phase. When keeping the proton number fixed and adding
neutrons to the system, the energy density
pattern slightly changes. The right panel of Fig. 2 corresponds
to the situation when the number of neutrons have been increased
to reach the total density $\rho=0.057$ fm$^{-3}$.
As shown in the right panel of Fig. 2,
the position of the minimum is not altered. It
suggests that the deformation of the ground state is mainly related
to the bound nucleons. On the other hand, there is a substantial
change of the position of the spherical phase which drops down
in energy and is now below the ``spaghetti'' phase.

The Fig. 2 shows that the change in the neutron density does not influence
qualitatively the structure of the ground state
since the energy minimum stays approximately at the same
proton quadrupole deformation.
However it is not always the case as one can see from Fig. 3,
where the energy density has been plotted as a function
of proton quadrupole moment for
larger total densities.
The left panel shows the results for $\rho=0.075$ fm$^{-3}$, and
the right one for $\rho=0.079$ fm$^{-3}$.
The two almost degenerate minima visible in the left panel
differ by the lattice type only, and both correspond
to nuclei immersed in a neutron gas.
The spherical configuration corresponds to a scc lattice and the deformed
one to a bcc lattice. There is also an excited
configuration associated with a slab-like (``lasagna'') phase.
Still increasing the number of neutrons, but at fixed proton number,
one sees on the right panel of Fig. 3, that the relative
position of the minima is not stable with respect to the
neutron density. The spherical phase is no longer energetically favored
and the bcc configuration  merges with the slab phase,
forming the lowest energy configuration in the form of ``rippled''
nuclear slabs.

The question arises whether these changes are related to the shell effects or
to the liquid drop part of the energy.
In order to better visualize this effect, we have plotted (see Fig.4) the
energy differences between different configurations:
\begin{itemize}
\item the spherical nuclear phase ($Q_{p}=0$ fm$^{2}$),
\item the ``spaghetti'' phase ($Q_{p}=2400$ fm$^{2}$),
\item the ``lasagna'' phase ($Q_{p}=2600$ fm$^{2}$).
\end{itemize}
The solid curve represents the quantity
$\Delta E/V = (E_{spherical}-E_{spaghetti})/V $ while
the dotted curve corresponds to
$\Delta E/V = (E_{spherical}-E_{lasagna})/V $.
One notices that these differences oscillate rapidly as a function of
the total density. It seems therefore
unlikely that the liquid drop part of the energy
is responsible for such rapid fluctuations.
On the other hand, the leading shell energy contribution to
the total energy, for two obstacles located at the distance $d$
and immersed in the neutron gas,
can be determined in the semiclassical approximation
(see appendix B):
\begin{equation} \label{shell}
E_{shell}^{out} \approx
\frac{\hbar^{2} L^{i} R^{2-i}}{8m_{n}} \left ( \frac{3}{\pi}
\right )^{\frac{2+i}{6}}
\frac{(\rho_{n}^{out})^{\frac{2+i}{6}}}{d^{\frac{6-i}{2}}}
\cos\left (2k^{n}_{F}d-i\frac{\pi}{4}\right ) ,
\end{equation}
where $m_{n}$ is the neutron
mass and $i=0,1,2$ stands for two spherical, cylindrical and planar obstacles,
respectively (we assume that rods and slabs are parallel to each other).
In the above equation, $L$ defines the length of the obstacle, and $R$ is
its radius (in the case of a slab it is defined
as half of its width).
This expression exhibits oscillations, where the gross staggering pattern
comes from the shortest periodic orbit of length $d$ between nearest nuclei.
This can be used to estimate qualitatively how the difference between the
shell energies of the two minima varies as a function of the
neutron density.

Assuming that the distance between the nearest nuclei is given by $r$, the
shell energy contribution due to unbound neutrons has the form:
\begin{equation}
E^{out}_{shell} \propto \cos(2k^{n}_{F} r).
\end{equation}
The quantity plotted in  Fig. 4 represents the difference between the energy
density  of two phases, and if it is associated with
shell effects, its behavior should be expressible as
a sum of cosine (or sine) functions. In particular, the leading
contribution with the longest period
comes from the single repetition of the shortest periodic orbits:
\begin{equation}
\Delta E^{out}_{shell} \approx
A_{i}\cos(2k^{n i}_{F}r_{i})+
A_{j}\cos(2k^{n j}_{F}r_{j}) \mbox{\hspace{0.5cm} $i\neq j$},
\end{equation}
where  $i,j=0,1,2$ for spherical nuclear phase, the
``spaghetti'' phase, and the ``lasagna'' phase, respectively.
The coefficient
$A_{i}$ is related to the stability of the orbit, $r_{i}$
is the distance between nuclear inhomogeneities, and
$k^{n i}_{F}$ stands for the Fermi momentum in the phase $i$.
One can infer from this expression that
the interference between the contributions of different orbits
is at the origin of the fluctuations of $\Delta E^{out}_{shell}$
as a function of the density.

Indeed, taking $k_{F}^{n i} = k^{n}_{F}
\approx (3\pi^2\rho^{out}_{n})^{1/3}$ and
assuming that  $A_{i}\approx A_{j}=A$ we get:
\begin{equation} \label{osc}
\Delta E^{out}_{shell}\approx
2A\sin(2k^{n}_{F}r_{ij})\sin(k^{n}_{F}\Delta r_{ij}),
\end{equation}
where $r_{ij}=\frac{1}{2}(r_{i}+r_{j})$ and $\Delta r_{ij}=(r_{i}-r_{j})$.
Hence the shell energy difference oscillates
as a function of the Fermi momentum
with a period $\Delta k^{n}_{F} =
\displaystyle{\frac{\pi}{r_{ij}}}$, or equivalently,
as a function of the neutron Fermi energy, with the period
$\Delta \epsilon^{n}_{F}=
\displaystyle{\frac{\hbar^{2}\pi^{2}}{2m_{n}r_{ij}^{2}}}$.

This quantity can be estimated from our HF calculations.
Let us define the radius of the nuclear inhomogeneity
as the distance between the center of a nucleus and the point
where the neutron density
decreases to the value $\frac{1}{2}(\rho^{out}_{n}+\rho^{in}_{n}) $.
This gives a radius $R_{0}\approx 6.23 fm$ for the spherical
and $R_{1}\approx 3.86 fm$ for the ``spaghetti'' phases,
at a density $\rho\approx 0.055$ $fm^{-3}$.
Since the length of the lattice $d$ is in this case
equal to $26 fm$, the distance between the two nearest spherical-like
and spaghetti-like objects can be estimated as
$r_{0}=d-2R_{0}\approx 13.54 fm$,
and $r_{1}=d-2R_{1}\approx 18.28 fm$, respectively.
This gives a period of oscillation
equal to $\Delta \epsilon^{n}_{F}\approx 0.8$ MeV. On the other hand, the
HF calculations predict the neutron Fermi energies to be equal to:
\begin{eqnarray}
\epsilon_{F}^{n}(Q_{p}=0,\rho=0.063)&=&12.043 MeV,   \nonumber \\
\epsilon_{F}^{n}(Q_{p}=0,\rho=0.053)&=&10.74 MeV, \nonumber \\
\epsilon_{F}^{n}(Q_{p}=2400,\rho=0.063)&=&11.931 MeV, \\
\epsilon_{F}^{n}(Q_{p}=2400,\rho=0.053)&=&10.607 MeV, \nonumber 
\end{eqnarray}
which give the periods:
\begin{eqnarray}
\Delta \epsilon^{n}_{F} (Q_{p}=0) &=& 1.303 MeV, \\
\Delta \epsilon^{n}_{F} (Q_{p}=2400) &=& 1.324 MeV.
\end{eqnarray}

They have to be compared with the $0.8$ MeV obtained
on the basis of the semiclassical analysis. The difference
between these values is mainly related
to the fact that in the semiclassical approximation we have neglected both
the coupling of the shell effects
to the liquid drop terms, and  higher order corrections
related to other orbits (including
the multiple repetitions of the shortest orbits). Nevertheless,
taking into account how drastic these approximations are, the agreement
is surprisingly good. It is even better for the case of the energy density
difference between the
spherical  and the slab phases (dotted curve in
the Fig. 4). The radii  are then: $R_{0}\approx 5.9 fm$,
and $R_{2}\approx 3.59 fm $ at a density $\rho\approx 0.075$
$fm^{-3}$, with the ``radius'' of the slab  defined
as half of its width. The distances between the inhomogeneities
for these two phases
are $r_{0}\approx 9 fm$ and $r_{2}\approx 13.62 fm$, respectively.
Half of the oscillation period
is equal to $\frac{1}{2}\Delta \epsilon^{n}_{F}\approx 0.801 MeV$.
On the other hand, according to the
HF calculations the neutron Fermi energies are equal to:
\begin{eqnarray}
\epsilon_{F}^{n}(Q_{p}=0,\rho=0.079)&=&12.645 MeV,   \nonumber \\
\epsilon_{F}^{n}(Q_{p}=0,\rho=0.0749)&=&11.779 MeV, \nonumber \\
\epsilon_{F}^{n}(Q_{p}=2600,\rho=0.079)&=&12.759 MeV, \\
\epsilon_{F}^{n}(Q_{p}=2600,\rho=0.0749)&=&11.884 MeV, \nonumber 
\end{eqnarray}
which give for half of the periods the values:
\begin{eqnarray}
\frac{1}{2}\Delta \epsilon^{n}_{F} (Q_{p}=0) &=& 0.866 MeV, \\
\frac{1}{2}\Delta \epsilon^{n}_{F} (Q_{p}=2400) &=& 0.875 MeV. \\
\end{eqnarray}
The agreement with the semiclassical estimation
is very good. This demonstrates that the oscillatory effect of the
energy density difference between the phases is due to the shell
energy of unbound neutrons. From eq. (\ref{osc}), it is clear
that another oscillatory effect may appear with a much longer period
$\tilde{\Delta} k^{n}_{F} =
\displaystyle{\frac{2\pi}{\Delta r_{ij}}}$.
Its mechanism is somewhat similar to the well known supershell phenomenon in
atomic clusters \cite{bbh}.
This effect however, requires the stability of one phase
over a wide density range and therefore may be not well
pronounced.

Up to now, we have neglected the pairing interaction.
The pairing energy is a smooth function of the density and is
not expected to play an important role for
the liquid drop part of the total energy expansion. On the contrary,
pairing correlations may
considerably influence the shell energy \cite{ybm,bcf}.

The pairing energy can be divided with a good approximation into
two parts. The first one comes from the pairing interaction
acting among bound nucleons, and the second one is associated
with pairing of unbound neutrons. These two contributions
can be, in the first approximation, treated as independent
quantities, because of the short range of the pairing interaction.
For example, in the usual zero-range approximation of the pairing interaction
(delta force) the amplitude for pair scattering is proportional
to the integral: $\displaystyle{\int
\phi_{k}^{*}({\bf r})
\phi_{l}^{*}({\bf r})
\phi_{m}({\bf r})
\phi_{n}({\bf r}) d^{3} r}$, where $\phi_{k}({\bf r})$ are
the single-particle wave functions. On the other hand,
bound single-particle states, as well as resonant states inside
nuclei have a small overlap with states forming continuum.
Hence the integral describing the scattering
of a Cooper pair from nuclei
to continuum or vice versa will be relatively small, and can
be, in a first approximation, neglected.

The pairing interaction acting among bound nucleons will
decrease the amplitude of shell effects inside the nuclei.
Note however that these effects can at most lead to a
slight deformation change of nuclei and do not influence the phase transition
pattern. Hence in the following we focus
only on the pairing energy associated with the neutron gas.
Since the pairing
gap $\Delta$ for neutron matter at subnuclear densities reaches
$3-4$ MeV at the Fermi level \cite{kkc}, the occupation of the
single-particle states close to the Fermi level
are significantly modified by pairing. In order to
analyze the influence of pairing on the shell structure, let us
consider a Fermi gas of neutrons in a volume $V$
at a density $\rho^{out}_{n}$, interacting via a pairing force.
The total energy
of the system can be expressed in the BCS approximation as:
\begin{equation}   \label{genpa}
E=\int_{-\infty}^{\infty}\epsilon v^{2}_{\mu}(\epsilon ) g(\epsilon )d\epsilon
+ E_{pair} ,
\end{equation}
where $g(\epsilon )$ is the density of single-particle states at the
energy $\epsilon$ and $\mu$ is the chemical potential defined by
the condition:
\begin{equation}
\rho_{n}^{out}=\frac{1}{V}
\int_{-\infty}^{\infty} v^{2}_{\mu}(\epsilon ) g(\epsilon )d\epsilon .
\end{equation}
The coefficient $v^{2}_{\mu}(\epsilon )$ represents the
BCS occupation factor:
\begin{equation} \label{bcsoc}
v^{2}_{\mu}(\epsilon )=\frac{1}{2}\left ( 1-
\frac{\epsilon-\mu}{\sqrt{ (\epsilon-\mu)^{2}+\Delta(\epsilon )^{2} }}
\right ) ,
\end{equation}
where $\Delta (\epsilon )$ is the (energy dependent) pairing gap.
In order to avoid divergences, we assume that the
function $\Delta$ approaches
zero sufficiently rapidly to make the integral in
equation (\ref{genpa}) convergent.
The term $E_{pair}$ denotes the pairing energy.
If the Fermi gas is subject to boundary conditions, e.g. due to
presence of impurities in the system,
the level density contains a fluctuating term:
\begin{equation}
g(\epsilon, {\bf d} )=\tilde{g}(\epsilon )+ g_{C}(\epsilon, {\bf d} ),
\end{equation}
where $\tilde{g}(\epsilon )$ is the smooth part
of the level density. The fluctuating part
$g_{C}(\epsilon, {\bf d} )$ depends on the shape and the mutual
geometrical arrangement of the impurities parametrized
by ${\bf d}$ and is responsible for the shell correction
energy. For two spherical, cylindrical or planar obstacles
located at a distance $d$, the semiclassical approximation gives
the leading contribution to $g_{C}$ in the form (see appendix B):
\begin{equation} \label{levden}
g_{C}(\epsilon, d )\approx \frac{1}{4} \frac{L^{i}R^{2-i}}{d^{1-i/2}}
\left (\frac{2m_{n}}{\pi^2\hbar^2}\right )^{\frac{i+2}{4}}
\epsilon^{\frac{i-2}{4}}\cos ( 2 k d-i\frac{\pi}{4} ) ,
\end{equation}
where $\epsilon=\displaystyle{\frac{\hbar^2 k^2}{2m_{n}}}$
and $i$ takes the values $0,1,2$  for the various shapes (compare
to eq. (\ref{shell})).

The total energy of the system can now be divided into
smooth and  oscillating parts (see appendix A):
\begin{eqnarray} \label{epair}
E&=&\tilde{E} + E_{C} + \Delta E + E_{pair} + O(1/V)  \nonumber \\
 &=&\int_{-\infty}^{\infty}\epsilon v_{\tilde{\mu}}^{2}(\epsilon )
\tilde{g}(\epsilon )d\epsilon +
\int_{-\infty}^{\infty}\epsilon v_{\tilde{\mu}}^{2}(\epsilon )
g_{C}(\epsilon, d )d\epsilon   \nonumber \\
 &-&
\int_{-\infty}^{\infty}v^{2}_{\tilde{\mu}}(\epsilon ) g_{C}(\epsilon, d)
d\epsilon \frac{\int_{-\infty}^{\infty}\epsilon \frac{\partial}
{\partial\tilde{\mu}}
v^{2}_{\tilde{\mu}}(\epsilon ) \tilde{g}(\epsilon)d\epsilon}
{\int_{-\infty}^{\infty} \frac{\partial}{\partial\tilde{\mu}}
v^{2}_{\tilde{\mu}}(\epsilon ) \tilde{g}(\epsilon)d\epsilon} + E_{pair}+
O(1/V),
\end{eqnarray}
where the first term represents the smooth part of the total energy.
The term $E_{C}$ is the Casimir energy, and
$\Delta E$ provides a correction to the Casimir energy due
to the Fermi energy renormalization in the presence of obstacles
\cite{bma1,bwi}.
The smooth part of the chemical potential
$\tilde{\mu}$ is calculated from the condition:
\begin{equation}
N=\int_{-\infty}^{\infty}v^{2}_{\tilde{\mu}}(\epsilon )
                         \tilde{g}(\epsilon )d\epsilon .
\end{equation}
It implies that the neutron
shell energy in the presence of pairing can be expressed in the
form:
\begin{equation}
E^{out}_{shell} = E_{C} + \Delta E + O(1/V),
\end{equation}
where we have neglected corrections due to the change of pairing
energy in the presence of obstacles (see appendix A).

In  Fig. 5 we  show the behavior of the shell energy as a function
of the distance between obstacles without pairing and with a pairing
interaction characterized by a pairing gap equal to
$\Delta=3.5$ MeV for $k < 20 fm^{-1}$
(a cutoff has been introduced to avoid
divergences). This pairing strength
corresponds to the pairing gap at the Fermi level
for a neutron Fermi momentum $k^{n}_{F} \approx 1$ fm$^{-1}$
obtained with the Argonne $v_{14}$ and $v_{18}$
interactions \cite{kkc}. One can see that
the amplitude of the shell energies are smaller but the decrease is
lower than 10\%, as
compared to the no pairing regime. We conclude from this comparison that
the pairing interaction will not
change qualitatively the HF results.

\section{CONCLUSIONS}

In this paper we have analyzed the structure of the neutron star
crust at zero temperature using the Hartree-Fock approach with
an effective Skyrme interaction.
In particular, we focused on the problem of shell effects
associated with unbound neutrons scattered by nuclear
inhomogeneities. Our results show that:
\begin{itemize}
\item the relative energies of different phases fluctuate rapidly
as a function of the total density,
\item the fluctuations can be attributed to the shell effects
associated with unbound neutrons,
\item the neutron shell energy density is of the same
order as the liquid drop energy density differences between the
various phases present in the crust, and its behavior
can be easily understood in terms of the periodic orbits
in the system,
\item the pairing correlations lead only
to a slight decrease of the shell
energy and are not expected to
influence significantly the structure of the crust.
\end{itemize}
Consequently, we may conclude that the
structure of the crust may be quite complicated since the shell
effects associated with  unbound neutrons
may easily reverse the phase transition order predicted by
liquid drop based approaches.  Moreover the number of phase
transitions may increase since the same phase may appear
for various density ranges (see Fig. 4).

Our results suggest also that several
phases, differing by the lattice type, could coexist in the crust.
This possibility was not taken into account in
previous investigations since in a liquid drop based approach,
the system favors only one lattice type for a given nuclear shape.
However, since the neutron shell
energy is very sensitive to the spatial order in the system,
this simple picture will change, and one may expect
the coexistence of various lattice geometries with different
lattice constants. Note also that
once a phase is formed, its geometric order will be stabilized by both the
Coulomb repulsion and the shell energy. Indeed,
although the Coulomb energy is a smooth function of the
nuclear displacement, the shell energy
is not. Since several different orbits contribute to the shell
effects (except for the slab--like phase) the displacement of a single
nucleus from its equilibrium position in the
lattice will give rise to interference effects
depending on the lattice type.
Hence it is likely that the system will favor distorted lattices, or lattices
with defects which decrease the shell energy.
In that respect the present results support the conclusions drawn on the
basis of a simple model \cite{bma1}.

All these arguments suggest that the inner crust is an extremely rich
and complicated system which may turn out to be on the verge
of a disordered phase.

\acknowledgements

This research was supported in part by the Polish Committee
for Scientific Research (KBN) under Contract No.~5~P03B~014~21
and the Wallonie/Brussels-Poland
integrated action program. Numerical calculations were performed
at the Interdisciplinary Centre for Mathematical and Computational
Modelling (ICM) at Warsaw University. Authors would like to thank
Hubert Flocard for providing us with the numerical code
solving the Poisson equation and
Aurel Bulgac for discussions and valuable comments.

\appendix
\section{Shell energy of the Fermi gas with impurities}

We derive here expressions for the shell energy
associated with impurities immersed in a fermionic
environment.

Let us consider a Fermi gas confined in a volume $V$,
large compared  to both the Fermi wavelength
($k_{F}^{3}V>>1$) and the volume of the obstacles.
At the end of the calculations the limit
$V\rightarrow \infty$ will be taken.
The shell energy is an
interaction energy between ``foreign''
objects immersed in the gas. Its origin comes from the
modification of the level density by the presence
of obstacles. The total level density can be decomposed
in the following way \cite{bbh,bwi}:
\begin{equation}
g(\epsilon, {\bf d})= \tilde{g} (\epsilon )
+ g_{C}(\epsilon, {\bf d}),
\end{equation}
where $\tilde{g}$ is the smooth part of $g$
and can be obtained either from the Weyl expansion
\cite{bbl,bbh} (in the case of impenetrable objects) or
within the Thomas-Fermi approach \cite{bbh}.
The set of parameters ${\bf d} = (d_{1}, d_{2}, ..., d_{n})$
describes the  geometrical arrangements of the obstacles.
The smooth part can also be treated as
the total density of states for a system with obstacles located at
infinity. Namely:
\begin{equation}
\lim_{d_{i}\rightarrow \infty} g(\epsilon, {\bf d}) = \tilde{g}(\epsilon),
\mbox{ \hspace{0.5cm} for i=1,2,...n}.
\end{equation}
The fluctuating part of the level density $g_{C}$
is associated with the formation of resonances. Its evaluation
in a semiclassical approximation is presented in the next section.

The shell energy can be defined in a form similar to the
Casimir energy \cite{bma1,bma2,bma3,bma4,bwi,pbh}:
\begin{equation} \label{eint}
E_{shell}({\bf d})=\int_{-\infty}^{\mu}
d\epsilon \epsilon g(\epsilon, {\bf d}) -
\int_{-\infty}^{\tilde{\mu}}d\epsilon \epsilon
\tilde{g}(\epsilon),
\end{equation}
where $\mu$ and $\tilde{\mu}$ denote the chemical potentials
associated with the total and smooth level densities, respectively.
They are related to each other through the
particle number conservation condition:
\begin{equation} \label{ncon}
N=\int_{-\infty}^{\mu}d\epsilon g(\epsilon, {\bf d}) =
\int_{-\infty}^{\tilde{\mu}}d\epsilon
\tilde{g}(\epsilon) .
\end{equation}
The decomposition of the integrals in equation (\ref{eint}) results
in the following expression:
\begin{equation}
E_{shell}({\bf d})=\Delta E + E_{C} =
\int_{\tilde{\mu}}^{\mu}
\epsilon \tilde{g}(\epsilon)d\epsilon +
\int_{-\infty}^{\mu}
\epsilon g_{C}(\epsilon, {\bf d})d\epsilon ,
\end{equation}
where $\Delta E$ represents the correction to the
shell energy due to the variation
of the chemical potential induced
by the presence of obstacles, while $E_{C}$ corresponds
to the ''pure Casimir--like'' force. Since
$\Delta\mu=\mu-\tilde{\mu}\propto O(1/V)$ and $\tilde{g}\propto V$
then
\begin{eqnarray}
\int_{\tilde{\mu}}^{\mu}
\epsilon\tilde{g}(\epsilon)d\epsilon &=&
\mu \Delta\mu \tilde{g}(\mu) + O( 1/V ), \nonumber \\
\int_{\tilde{\mu}}^{\mu}
\tilde{g}(\epsilon)d\epsilon &=&
\Delta\mu \tilde{g}(\mu) + O( 1/V ) ,
\end{eqnarray}
and thus it is clear that both $\Delta E$ and $E_{C}$
do not vanish in the limit of infinite volume.
The condition (\ref{ncon}) can be used to calculate $\Delta\mu$:
\begin{equation}
\Delta\mu=
-\frac{1}{\tilde{g}(\mu)}\int_{0}^{\mu}g_{C}(\epsilon, {\bf d})
d\epsilon + O(1/V^2 ).
\end{equation}

The shell energy can be finally written in the form:
\begin{equation}
E_{shell}({\bf d})=\int_{-\infty}^{\mu}
( \epsilon - \mu )g_{C}(\epsilon, {\bf d})d\epsilon + O( 1/V ).
\end{equation}
Taking the limit $V\rightarrow \infty$, the higher
order terms vanish and the first term in the above
formula yields an exact expression for the shell energy.

In the case of a superfluid  Fermi gas, the energy gap depends
on the momentum $k$. To avoid divergences in
the expression for the total energy, one has to  assume
that $\Delta (k)$ is tending to zero at least like $1/k$
for $k\rightarrow \infty$.
The pairing properties of the
system are affected by the presence of obstacles and thus the
solution of the gap equation with impurities $\Delta(k, {\bf d})$
differs slightly from the smooth pairing gap
$\tilde{\Delta}(k)$ obtained using the smooth level density:
$\Delta (k, {\bf d})
=\tilde{\Delta} (k) + \Delta_{C}(k, {\bf d})$. $\Delta_{C}$ denotes
the quantum correction to the smooth pairing gap and is due to the fact that
the level density is modified. Hence, eq. (\ref{eint})
is generalized for  superfluid systems in the form:
\begin{eqnarray} \label{einpair}
E_{shell}({\bf d})&=&\int_{-\infty}^{\infty}
d\epsilon \epsilon v^{2}_{\mu}(\epsilon, {\bf d})g(\epsilon, {\bf d}) -
\int_{-\infty}^{\infty} d\epsilon \epsilon
\tilde{v}^{2}_{\tilde{\mu}}(\epsilon)\tilde{g}(\epsilon)
\nonumber \\
&+& E_{pair}({\bf d}) - \tilde{E}_{pair},
\end{eqnarray}
where $v^{2}_{\mu}$ is the BCS occupation factor
in the presence of obstacles (\ref{bcsoc}) and
the term $\tilde{v}_{\tilde{\mu}}$ represents the smooth
occupation factor obtained for impurities far apart from each other.
The last two terms  are due to the correction of
the pairing energy related to the appearance of
resonant states between obstacles.

For simplicity, we assume  that $\Delta_{C}=0$
(and thus $\tilde{v}^{2}_{\mu} = v^{2}_{\mu}$) and
neglect the effect of the geometrical arrangements
of the obstacles on the pairing properties of the system.
The  equation simplifies then to the form:
\begin{equation} \label{einpair1}
E_{shell}({\bf d})=\int_{-\infty}^{\infty}
d\epsilon \epsilon
v^{2}_{\mu}(\epsilon)g(\epsilon, {\bf d}) -
\int_{-\infty}^{\infty} d\epsilon \epsilon
v^{2}_{\tilde{\mu}}(\epsilon)\tilde{g}(\epsilon),
\end{equation}
where the chemical potentials $\mu$ and $\tilde{\mu}$
are determined from the conditions:
\begin{equation} \label{npair}
N=\int_{-\infty}^{\infty}d\epsilon
v^{2}_{\mu}(\epsilon)g(\epsilon, {\bf d}) =
\int_{-\infty}^{\infty}d\epsilon
v^{2}_{\tilde{\mu}}(\epsilon)\tilde{g}(\epsilon) .
\end{equation}
Expanding $v^{2}_{\mu}$ around $\tilde{\mu}$ one gets:
\begin{equation}
E_{shell}({\bf d})=
\Delta\mu\int_{-\infty}^{\infty} d\epsilon \epsilon
\frac{\partial v^{2}_{\tilde{\mu}}(\epsilon)}
{\partial \tilde{\mu}} \tilde{g}(\epsilon) +
\int_{-\infty}^{\infty} d\epsilon \epsilon
\left (v^{2}_{\tilde{\mu}}(\epsilon)+ \Delta\mu
\frac{\partial v^{2}_{\tilde{\mu}}(\epsilon)}
{\partial \tilde{\mu}} \right ) g_{C}(\epsilon, {\bf d}) + O( 1/V ).
\end{equation}

Note however that the term:
$\displaystyle{\Delta{\mu}\int_{-\infty}^{\infty} d\epsilon \epsilon
\frac{\partial v^{2}_{\tilde{\mu}}(\epsilon)}
{\partial \tilde{\mu}} g_{C}(\epsilon, {\bf d})}$
is also proportional to $1/V$ and thus vanishes when
$V\rightarrow\infty$. Determining $\Delta\mu$ from condition
(\ref{npair}) one gets the expression for the shell energy:
\begin{equation} \label{einpair2}
E_{shell}({\bf d})=
\int_{-\infty}^{\infty} \epsilon
v^{2}_{\tilde{\mu}}(\epsilon ) g_{C}(\epsilon, {\bf d})d\epsilon
-\int_{-\infty}^{\infty}v^{2}_{\tilde{\mu}}(\epsilon ) g_{C}(\epsilon, {\bf d})
d\epsilon \frac{\int_{-\infty}^{\infty}\epsilon \frac{\partial}
{\partial\tilde{\mu}}
v^{2}_{\tilde{\mu}}(\epsilon ) \tilde{g}(\epsilon)d\epsilon}
{\int_{-\infty}^{\infty} \frac{\partial}{\partial\tilde{\mu}}
v^{2}_{\tilde{\mu}}(\epsilon ) \tilde{g}(\epsilon)d\epsilon} + O(1/V).
\end{equation}
Since $v^{2}_{\tilde{\mu}}(\epsilon)
\rightarrow \theta(\tilde{\mu}-\epsilon)$ and
$\displaystyle{
\frac{\partial v^{2}_{\tilde{\mu}}(\epsilon)}
{\partial \tilde{\mu}}}\rightarrow
-\delta(\epsilon-\tilde{\mu})$ when $\Delta(k)$ tends to zero,
the equation (\ref{einpair2}) becomes identical with (\ref{eint})
in the no pairing limit.

\section{Evaluation of the fluctuating part of level density
and the shell energy in the semiclassical approximation}

We derive here expressions for the shell (interaction)
energy for two identical, impenetrable obstacles at large separations
(\ref{shell}).  Three geometries of the obstacles are considered:
spherical, rod--like and planar geometries.
The system is characterized by a set
of parameters $R, L, d$ which denote the radius,
length (in the case of rods and slabs), and the distance between obstacles,
respectively. The distance is measured as half of the length of
the shortest periodic orbit. It means that e.g. in the case of
spheres the distance between their centers is equal to $d+2R$.
In the case of a slab, $R$ denotes half of its width.

In the semiclassical approximation, the fluctuating part of
the level density $g_{C}$ can be obtained
through the summation of the contributions associated
with the classically allowed periodic orbits in the system.
Recent exact calculations based on the S-matrix approach\cite{bwi}
indicated that this approximation
work surprisingly well even if the obstacles are close
to each other, i.e. when $d<<R$.
In general, the summation is a tedious task if many orbits contribute to the
level density. Especially when  elliptic or parabolic orbits are considered,
the problem requires a more careful treatment in order to avoid divergences
\cite{bbh}.  The semiclassical approximation to the interaction energy between
two obstacles is particularly useful if the periodic orbits of the system are
unstable (hyperbolic) and isolated \cite{bbh}. It holds true for  spherical
obstacles. For two spheres of radius $R$, located
at a distance $d$, the contribution to the level density arising
from the single periodic orbit is:
\begin{equation} 
g^{sphere}_{C}(k,d) = \frac{m d}{\pi\hbar^{2} k}\sum_{n=1}^{\infty}
\frac{1}{2 \sinh^{2}(\frac{n\lambda}{2})}\cos (2nkd-\sigma \frac{\pi}{2}),
\end{equation}
where $\lambda=
\displaystyle{2\log
\left (1+\frac{d}{R}+\sqrt{\frac{d}{R}\left (\frac{d}{R}+2\right )}
\right )}$
is the Lyapunov exponent determining the stability of the orbit,
$\sigma$ is the Maslov index depending on the type
of boundary conditions, and
$\epsilon=\displaystyle{\frac{\hbar^{2}k^{2}}{2m}}$.
For Dirichlet boundary
conditions, which we assume here, each reflection changes
the phase by $\pi$, thus
in our case $\sigma=4 n$ where $n$ is the number of repetitions of periodic
orbit.  The shell energy calculated from the expression (\ref{eint})
gives \cite{bma1}:
\begin{equation} \label{bub}
E^{sphere}_{shell}(d) = \frac{\hbar^2}{16\pi m d^2}\sum_{n=1}^{\infty}
\frac{1}{n^3 \sinh^{2}(\frac{n\lambda}{2})}
\left ( 2nk_{F}d \cos(2nk_{F}d) - \sin(2nk_{F}d)\right ).
\end{equation}
At large distances, i.e. for $\displaystyle{\frac{d}{R}}>>1$,
the sum can be truncated at $n=1$ and the expression simplifies to:
\begin{equation}  \label{esphere}
E^{sphere}_{shell}(d)\approx \frac{ \hbar^2 R^{2} }{ 8 m}
\left (\frac{3}{\pi}\right )^{1/3}\frac{\rho^{1/3}}{d^3}
\cos(2k_{F} d) + O(1/d^5),
\end{equation}
where we have expressed the amplitude of the shell energy in terms of
the density $\rho=
\displaystyle{\frac{k_{F}^{3}}{3\pi^2}}$.
This expression is
associated with the fluctuating level density:
\begin{equation} \label{gsphere}
g^{sphere}_{C}(k,d)
\approx\frac{1}{2}\frac{m R^2}{\pi\hbar^2 k d}\cos(2kd)
 + O(1/d^3).
\end{equation}

In the case of two parallel rods of length $L$, the orbit is no longer
isolated. Hence it is convenient to consider first the two-dimensional
case of two disks. Then the single periodic
orbit gives rise to the fluctuating part of the level density:
\begin{equation}
g^{disk}_{C}(k,d) = \frac{m d}{\pi\hbar^{2} k}\sum_{n=1}^{\infty}
\frac{1}{\sinh (\frac{n\lambda}{2})}\cos (2nkd).
\end{equation}
In order to obtain the $g^{rod}_{C}$ one has to calculate
first the fluctuating part of the particle number
$N^{rod}_{C}(\mu ,d)=
\displaystyle{\int_{-\infty}^{\mu}g^{rod}_{C}
(\epsilon, d) d\epsilon}$. But $N_{C}^{rod}$ has to be
an integral of the product of $g^{disk}_{C}$
and the level density of one-dimensional Fermi gas.
Namely, the quantity $N_{C}^{rod}$ can be expressed as:
\begin{eqnarray}
N^{rod}_{C}(k_{F},d)&=&\frac{L}{2\pi}\int_{0}^{k_{F}}g^{disk}_{C}(k)
d\left (\frac{\hbar^2 k^2}{2m}\right )
\int_{-\sqrt{k_{F}^2-k^{2}}}^{\sqrt{k_{F}^{2}-k^{2}}} dk_{p} \nonumber \\
&=&\frac{Lk_{F}}{4\pi}\sum_{n=1}^{\infty}
\frac{1}{n\sinh(\frac{n\lambda}{2})}J_{1}(2nk_{F}d),
\end{eqnarray}
where $k_{p}$ denotes the momenta parallel to rods and
$J_{1}$ is the Bessel function of the first kind.
Since $g^{rod}_{C}(k,d)=
\displaystyle{\frac{dN^{rod}_{C}}{dk}}\frac{m}{\hbar^2 k}$ we
get\footnote{In the Refs. \cite{bma1,bma2} the factor ``$1/\pi$'' has been
missed in the expression for the
shell energy and level density of the rod--like phase.}:
\begin{equation}
g^{rod}_{C}(k,d)=\frac{m L}{4\pi\hbar^{2} k}\sum_{n=1}^{\infty}
\frac{1}{n\sinh(\frac{n\lambda}{2})}
\left ( k d J_{0}(2nkd)+\frac{1}{n}J_{1}(2nkd)-kdJ_{2}(2nkd)\right ),
\end{equation}
and the shell energy:
\begin{equation}
E^{rod}_{shell}(d)=-\frac{\hbar^2 k_{F}^2}{8m\pi}\frac{L}{d}
\sum_{n=1}^{\infty}
\frac{1}{n^{2}\sinh(\frac{n\lambda}{2})}J_{2}(2nk_{F}d).
\end{equation}
In the limit $\displaystyle{\frac{d}{R}}>>1$, using the asymptotic
formula for a Bessel function and expressing the Fermi momentum
as a function of the density, one gets:
\begin{equation}    \label{grod}
g^{rod}_{C}(k,d)\approx\frac{1}{2}\frac{m L R}{\sqrt{kd\pi^3}\hbar^2}
\cos(2kd-\frac{\pi}{4})
 + O(1/d^{3/2}),
\end{equation}
and for the shell energy:
\begin{equation} \label{erod}
E^{rod}_{shell}(d)
\approx\frac{\hbar^2 L R}{8m}\left (\frac{3}{\pi}\right )^{1/2}
\frac{\rho^{1/2}}{d^{5/2}}\cos(2nk_{F}d-\frac{\pi}{4}) + O(1/d^{7/2}).
\end{equation}

In the case of two slabs the semiclassical approximation is no longer
applicable since the classical orbits in this case are parabolic
which implies that $\lambda=0$ and the expression for $g_{C}$
diverges. On the other hand, the problem can be
solved analytically  yielding an exact solution
for the shell energy \cite{bma1}. In the following, however
we  derive an expression in a different way,
which has the advantage that
the final formula has a form similar  to the semiclassical
expressions for spheres and rods.

We consider first the fluctuating level density of a
one-dimensional Fermi gas confined in a box of length $d$
with Dirichlet boundary conditions.
The total level density in this case can be calculated exactly
and has the form \cite{bbh}:
\begin{equation}
g(k,d)=\frac{m d}{\hbar^2 k\pi}( 1 + 2\sum_{n=1}^{\infty}\cos(2nkd) ).
\end{equation}
The second term corresponds to the fluctuating part of
the level density:
\begin{equation}
g^{1 dim}_{C}(k,d)=\frac{2 m d}{\hbar^2 k\pi}\sum_{n=1}^{\infty}\cos(2nkd) .
\end{equation}
Using an analogous approach as in the case of  rods, one
can calculate the fluctuating part of the particle number
as a function of the momentum:
\begin{eqnarray}
N^{slab}_{C}(k_{F})&=&\left (\frac{L}{2\pi}\right )^2
\int_{0}^{k_{F}}g^{1 dim}_{C}(k_{z},d)
d\left (\frac{\hbar^2 k_{z}^2}{2m}\right )
\int_{k_{x}^2+k_{y}^2<k_{F}^{2}-k_{z}^{2}} dk_{x}dk_{y} \nonumber \\
&=&\left (\frac{L}{2\pi}\right )^2\frac{1}{2d^{2}}
\sum_{n=1}^{\infty}\frac{-2nk_{F}d\cos(2nk_{F}d)+\sin(2nk_{F}d)}{n^3},
\end{eqnarray}
where $k_{z}$ is the momentum component perpendicular to
slabs. Hence the fluctuating level density for slabs is given by:
\begin{equation}
g^{slab}_{C}(k,d)=\frac{dN^{slab}_{C}}{dk}\frac{m}{\hbar^{2}k}=
2m\left (\frac{L}{2\pi\hbar}\right )^2
\sum_{n=1}^{\infty}\frac{\sin(2 n k d)}{n}.
\end{equation}
The shell energy calculated using
expression (\ref{eint}) gives:
\begin{equation}
E^{slab}_{shell}=\frac{\hbar^2 L^2 d}{32 m \pi^2}
\sum_{n=1}^{\infty}\frac{1}{(nd)^5}
(6 nk_{F}d\cos(2nk_{F}d) + 4(nk_{F}d)^{2}\sin(2nk_{F}d)
-3\sin(2nk_{F}d) ).
\end{equation}
Note that unlike in the previous cases the truncation of the summation
for large separation is not justified since all terms
have the same  dependence on the distance. Nevertheless, the gross shell
structure is still represented by the $n=1$-term and thus
for comparison with previous expressions we
consider the leading term $n=1$:
\begin{equation}    \label{eslab}
E^{slab}_{shell} (d)\approx\frac{\hbar^2 L^2}{8 m}
\left (\frac{9}{\pi^2}\right )^{1/3}\frac{\rho^{2/3}}{d^2}
\sin(2k_{F}d) + O(1/d^{2}),
\end{equation}
and for the fluctuating level density:
\begin{equation}  \label{gslab}
g_{C}^{slab}(k,d)\approx
\frac{1}{2}\frac{m L^2}{\pi^2 \hbar^2}\cos (2kd-\frac{\pi}{2})
 + O(1/d^{0}).
\end{equation}

The expressions (\ref{esphere},\ref{erod},
\ref{eslab}) and (\ref{gsphere},\ref{grod},\ref{gslab})
can be gathered into a single formula for the shell energy and the
fluctuating part of the level density, respectively:
\begin{equation}
E^{i}_{shell} (d)\approx
\frac{\hbar^{2} L^{i} R^{2-i}}{8m} \left ( \frac{3}{\pi}
\right )^{\frac{2+i}{6}}
\frac{(\rho)^{\frac{2+i}{6}}}{d^{\frac{6-i}{2}}}
\cos\left (2k_{F}d-i\frac{\pi}{4}\right )
+ O(1/d^{5-\frac{3i}{2}}),
\end{equation}
\begin{equation}
g^{i}_{C}(\epsilon, d )\approx \frac{1}{4} \frac{L^{i}R^{2-i}}{d^{1-i/2}}
\left (\frac{2m}{\pi^2\hbar^2}\right )^{\frac{i+2}{4}}
\epsilon^{\frac{i-2}{4}}\cos\left ( 2 k d-i\frac{\pi}{4} \right )
+ O(1/d^{3-\frac{3i}{2}}),
\end{equation}
where $i=0,1,2$ correspond to the spherical, rod and slab case, respectively
and $\epsilon=\displaystyle{\frac{\hbar^2 k^{2}}{2m}}$.

\newpage

\begin{figure}[ht]
\caption[FIG1]{%
The total energy density (\ref{energy})
of the npe matter confined
within the cubic box of length $d$ as a function
of the proton fraction. The SLy4 force has been used
for the nucleon-nucleon interaction. All densities
are specified in fm$^{-3}$.
 } \label{FIG1}
\end{figure}

\begin{figure}[ht]
\caption[FIG2]{%
The total energy density (\ref{energy}) of
the npe matter as a function of the proton
quadrupole moment $Q_p=Q^{p}_{20}$ (middle subfigures).
The integrated proton and neutron densities (see text
for definition) corresponding
to nuclear configurations indicated by arrows are shown
in the lower and upper subfigures.
}
\label{FIG2}
\end{figure}

\begin{figure}[ht]
\caption[FIG3]{%
The same as in the Fig.2 but for different densities.
}
\label{FIG3}
\end{figure}

\begin{figure}[ht]
\caption[FIG4]{%
The energy density difference $\Delta E/V$ between
nuclear phases as a function of the total density.
Solid curve denotes the difference
between the spherical and rod-like phase.
Dotted curve denotes the difference
between the spherical and slab-like phase.
Smaller subfigures show the energy density of the npe
matter as a function of the proton quadrupole moment
for four different densities. Parameter $d$ denotes
the length of the cubic box.
}
\label{FIG4}
\end{figure}

\begin{figure}[ht]
\caption[FIG5]{%
The gross structure of the interaction energy (shell energy)
as a function of a distance
between two impenetrable obstacles immersed in the neutron gas
at the density $\rho^{out}_{n}=0.05$ fm$^{-3}$, calculated
in the semiclassical approximation.
The radius of the spherical obstacle (a) $R=7$ fm,
the radius and length of the cylindrical obstacle (b) is equal
to $R=4$ fm and $L=28.58$ fm, respectively.
The width and length of the planar obstacle (c) is equal to $R=7$ fm
and $L=14.32$ fm. The sizes were adjusted to keep
the volume of the obstacles constant.
}
\label{FIG5}
\end{figure}


\begin{thebibliography}{99}
\bibitem{bbp} G. Baym, H.A. Bethe, C.J. Pethick, Nucl. Phys.
              {\bf A175}, 225 (1971).
\bibitem{bba} J.-R. Buchler, Z. Barkat, Phys. Rev. Lett. {\bf 27},
              48 (1971).
\bibitem{rbp} D.G. Ravenhall, C.D. Bennett, C.J. Pethick,
              Phys. Rev. Lett. {\bf 28}, 978 (1972).
\bibitem{arp} J. Arponen, Nucl. Phys. {\bf A191}, 257 (1972).
\bibitem{nvb} J.W. Negele and D. Vautherin, Nucl. Phys. {\bf A207},
              298 (1973); P. Bonche and D. Vautherin, Nucl. Phys.
             {\bf A372}, 496 (1981); Astron. Astrophys. {\bf 112}, 268 (1982).
\bibitem{rpw} D.G. Ravenhall, C.J. Pethick, J.R. Wilson,
              Phys. Rev. Lett. {\bf 50}, 2066 (1983).
\bibitem{hsy} M. Hashimoto, H. Seki, M. Yamada, Prog. Theor. Phys.
              {\bf 71}, 320 (1984).
\bibitem{ohy} K. Oyamatsu, M. Hashimoto, M. Yamada, Prog. Theor. Phys.
              {\bf 72}, 373 (1984).
\bibitem{lpr} J.M. Lattimer, C.J. Pethick, D.G. Ravenhall,
              Nucl. Phys. {\bf A432}, 646 (1985).
\bibitem{wko} R.D. Wilson and S.E. Koonin, Nucl. Phys. {\bf A435}, 844
              (1985).
\bibitem{lfb} M. Lassaut, H. Flocard, P. Bonche, P.-H. Heenen, E. Suraud,
              Astron. Astrophys. {\bf 183}, L3 (1987).
\bibitem{oya1} K. Oyamatsu, Nucl. Phys. {\bf A561}, 431 (1993).
\bibitem{lor} C.P. Lorenz, D.G. Ravenhall, C.J. Pethick,
              Phys. Rev. Lett. {\bf 70}, 379 (1993).
\bibitem{oya2} K. Oyamatsu, M. Yamada, Nucl. Phys. {\bf A578}, 181 (1994).
\bibitem{pra} C.J. Pethick and D.G. Ravenhall, Annu. Rev. Nucl. Part. Sci.
              {\bf 45}, 429 (1995).
\bibitem{dhm} F. Douchin, P. Haensel, J. Meyer,
              Nucl. Phys. {\bf A 665}, 419 (2000).
\bibitem{dha} F. Douchin, P. Haensel, Phys. Lett. {\bf B485}, 107 (2000).
\bibitem{bma1} A. Bulgac, P. Magierski, Nucl. Phys. {\bf A683}, 695 (2001).
\bibitem{bma2} A. Bulgac, P. Magierski, Phys. Scripta {\bf T90}, 150 (2001).
\bibitem{bma3} A. Bulgac, P. Magierski, Acta Phys. Pol. {\bf B32}, 1099 (2001).
\bibitem{cas} H.B.G. Casimir, Proc. K. Ned. Akad. Wet. {\bf 51}, 793
              (1948); V.M. Mostepanenko and N.N. Trunov, Sov. Phys. Usp.
              {\bf 31}, 965 (1988) and references therein.
\bibitem{fish} M.E. Fisher and P.G. de Gennes, C.R. Acad. Sci. Ser. B
              {\bf 287}, 207 (1978); A. Hanke {\it et al.},
              Phys. Rev. Lett. {\bf {81}}, 1885 (1998) and references therein.
\bibitem{kgo} M. Kardar and R. Golestanian, Rev. Mod. Phys.
              {\bf 71}, 1233 (1999) and references therein.
\bibitem{bma4} P. Magierski, A. Bulgac, Acta Phys. Pol. {\bf B32} 2713 (2001).
\bibitem{bwi} A. Bulgac, A. Wirzba, Phys. Rev. Lett. {\bf 87}, 120404 (2001).
\bibitem{rki} M.A. Ruderman, C. Kittel, Phys. Rev. {\bf 96}, 99 (1954).
\bibitem{fwa} A.L. Fetter and J.D. Walecka, {\it Quantum Theory of Many
              Particle Systems}, (McGraw--Hill, New York, 1971).
\bibitem{bfh} P. Bonche, H. Flocard, P.-H. Heenen, S.J. Krieger, M.S. Weiss,
              Nucl. Phys. {\bf A443}, 39 (1985).
\bibitem{dfk} K.T.R. Davies, H. Flocard, S. Krieger, M.S. Weiss,
              Nucl. Phys. {\bf A443} 39 (1985).
\bibitem{cmb}
 E. Chabanat, P. Bonche, P. Haensel, J. Meyer and R. Schaeffer,
             Nucl.~Phys.~{\bf A635} (1998) 231.
\bibitem{wff} R.B. Wiringa, V. Fiks, A. Fabrocini, Phys. Rev. {\bf 38},
              1010 (1988).
\bibitem{pbh} P. Magierski, A. Bulgac, P.-H. Heenen,
              Int. J. Mod. Phys. {\bf A} (submitted).
\bibitem{bbh} M. Brack and R.K. Bhaduri, {\it Semiclassical Physics},
              Addison--Wesley, Reading, MA (1997).
\bibitem{ybm} Y. Yu, A. Bulgac, P. Magierski, Phys. Rev. Lett. {\bf 84}, 412
              (2000).
\bibitem{bcf} A. Bulgac, S.A. Chin, H.A. Forbert, P. Magierski, Y. Yu
              in Proc. of the Int. Workshop on {\it Collective excitations
              in Fermi and Bose systems''},
              eds. C.A. Bertulani, L.F. Canto and M.S. Hussein, pp. 44--61,
              World Scientific, Singapore (1999); nucl-th/9811028.
\bibitem{kkc} V.A. Khodel, V.V. Khodel, J.W. Clark, Nucl. Phys. {\bf A598},
              390 (1996).
\bibitem{bbl} R. Balian, C. Bloch, Ann. Phys. (N.Y.) {\bf 60}, 401 (1970);
              {\bf 63}, 592, (1971); {\bf 64}, 271 (1971); Erratum: {\bf 84},
               559 (1974).

\end{thebibliography}
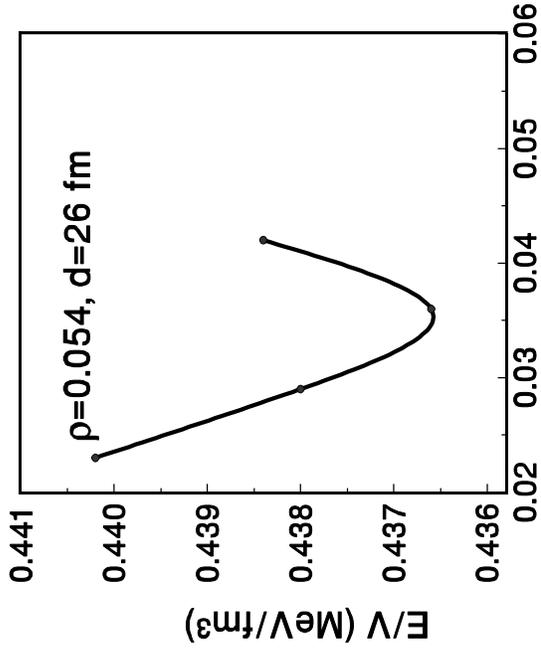
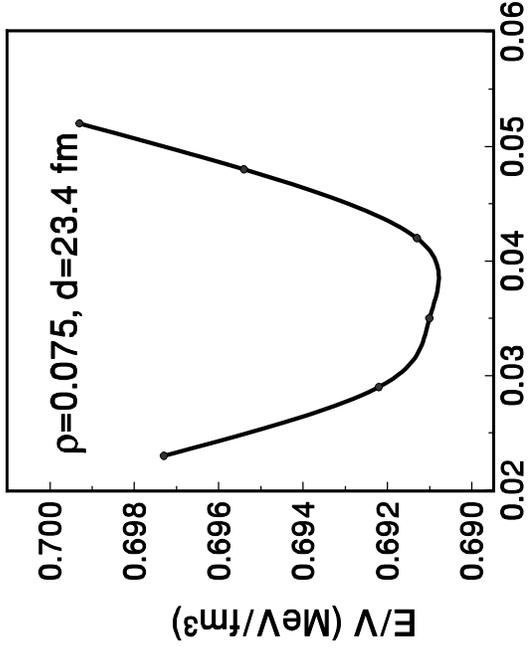
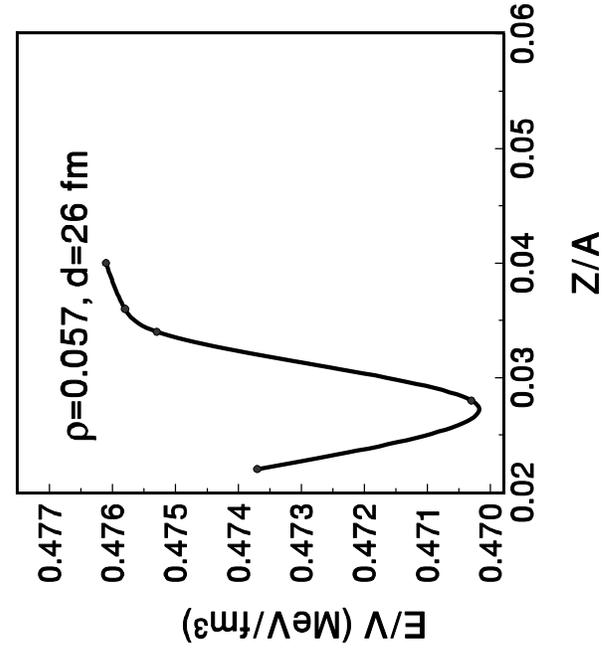
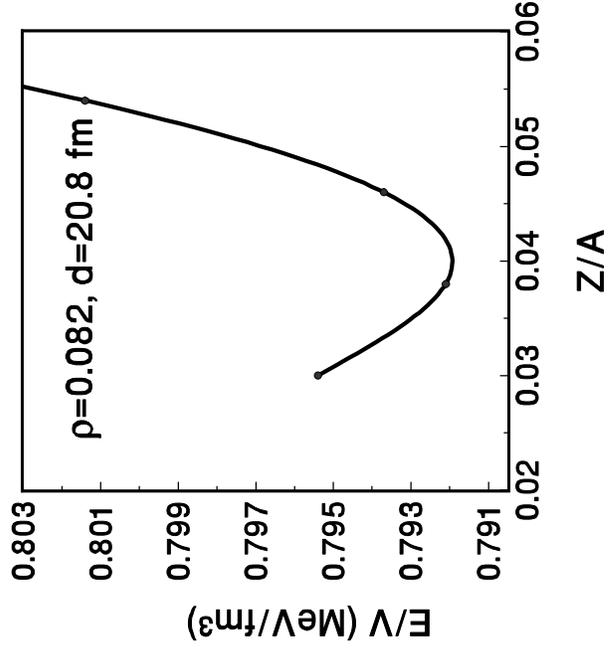
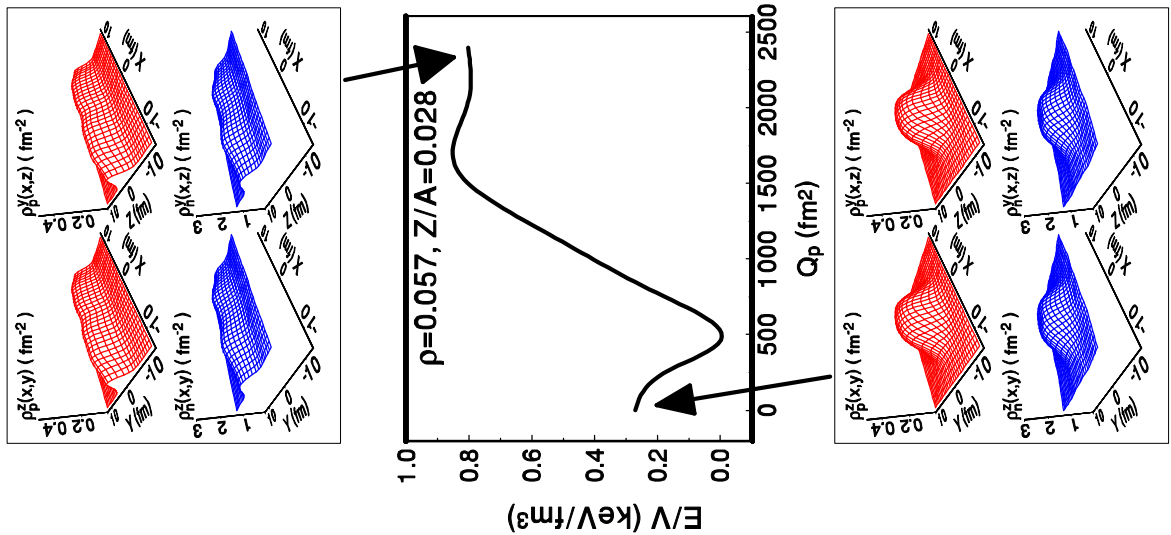
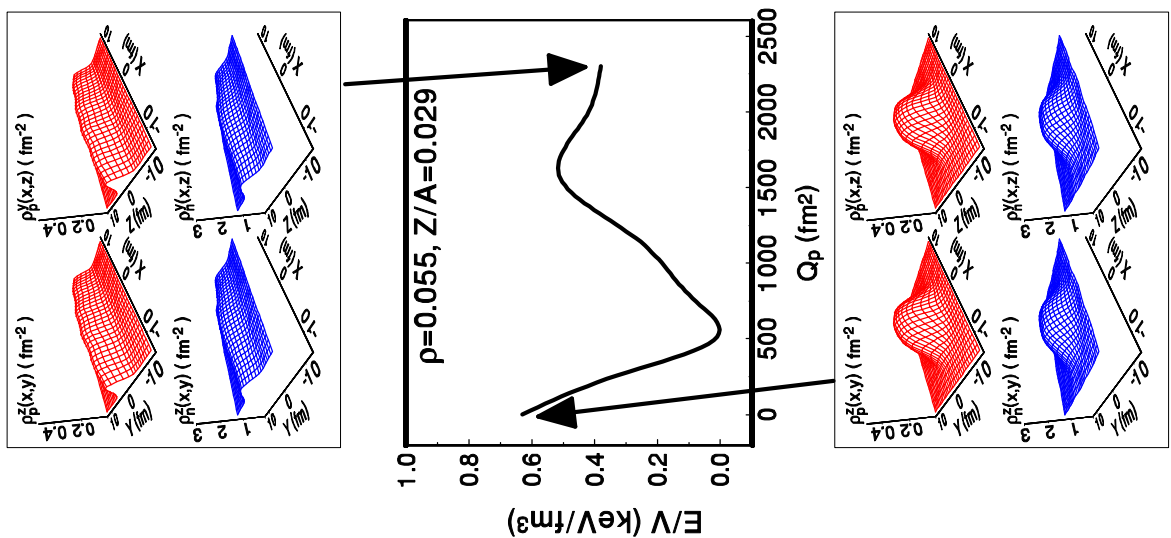
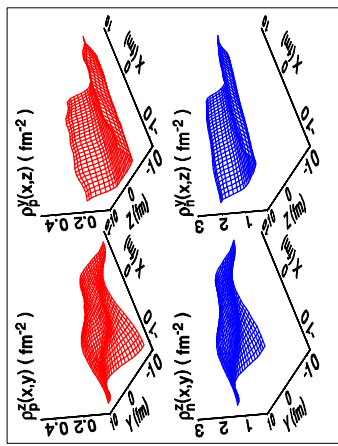
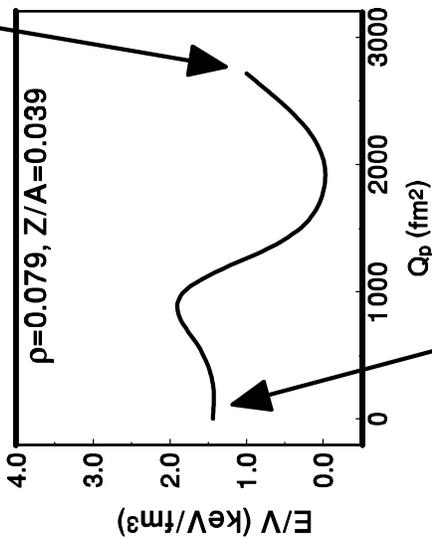
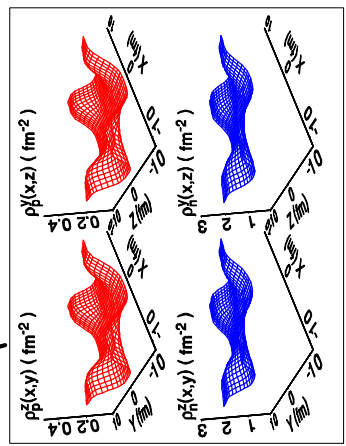
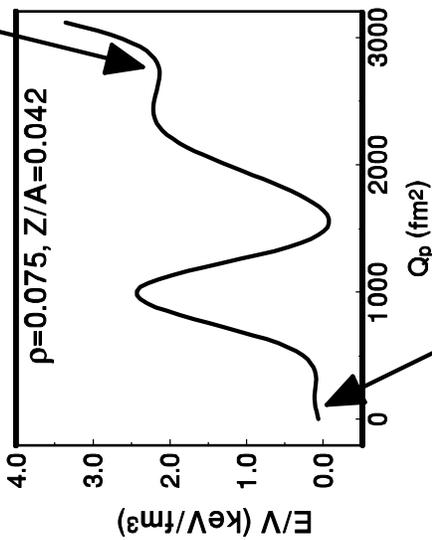
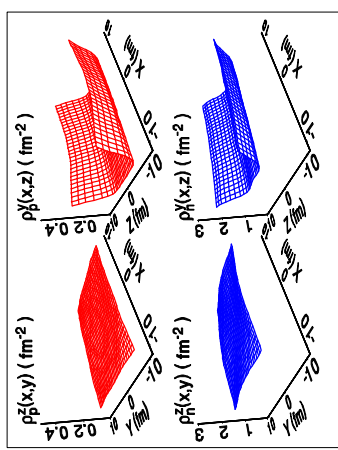
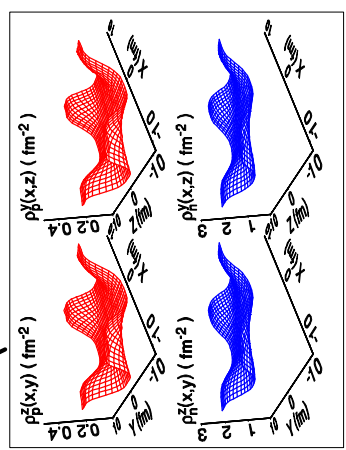
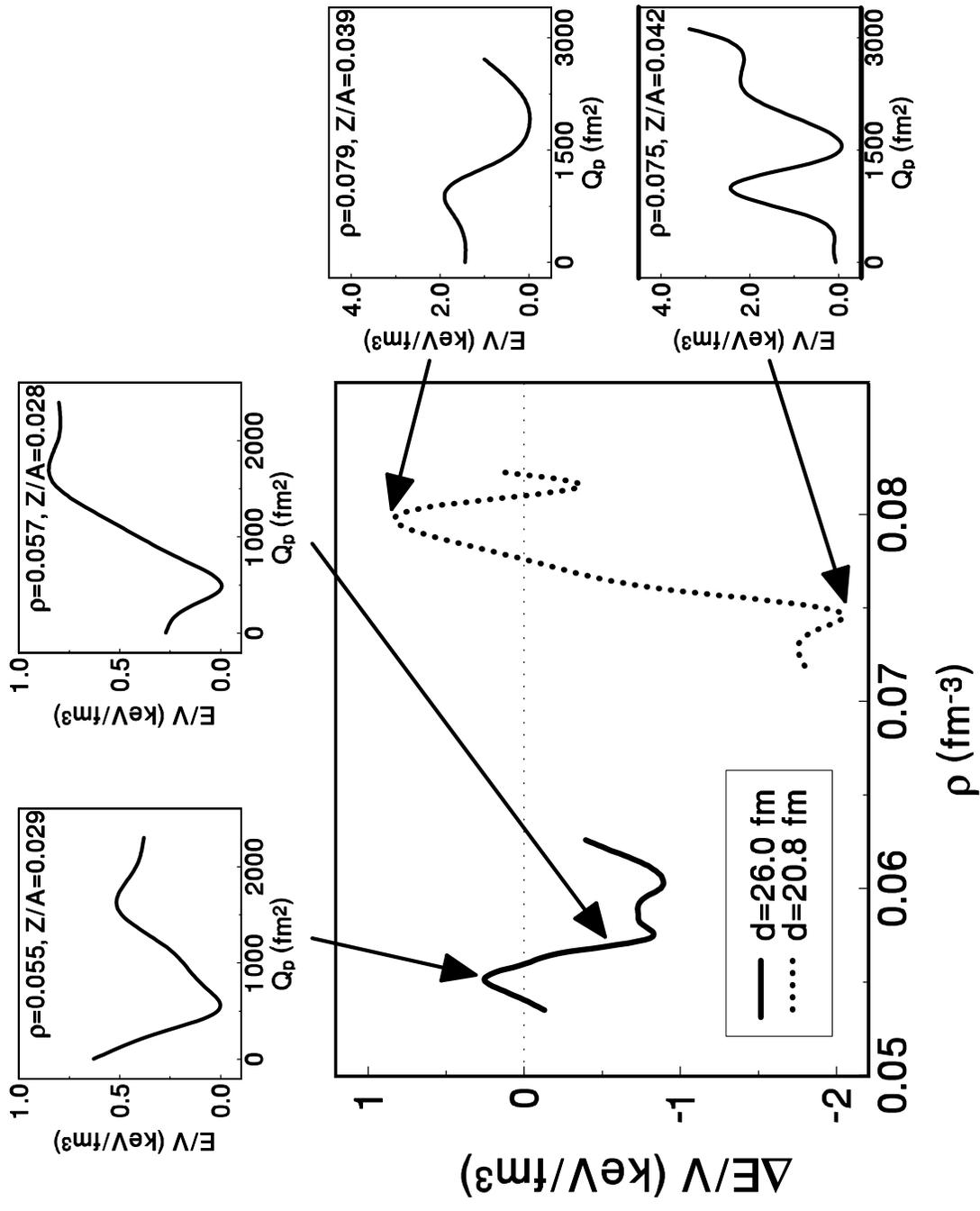
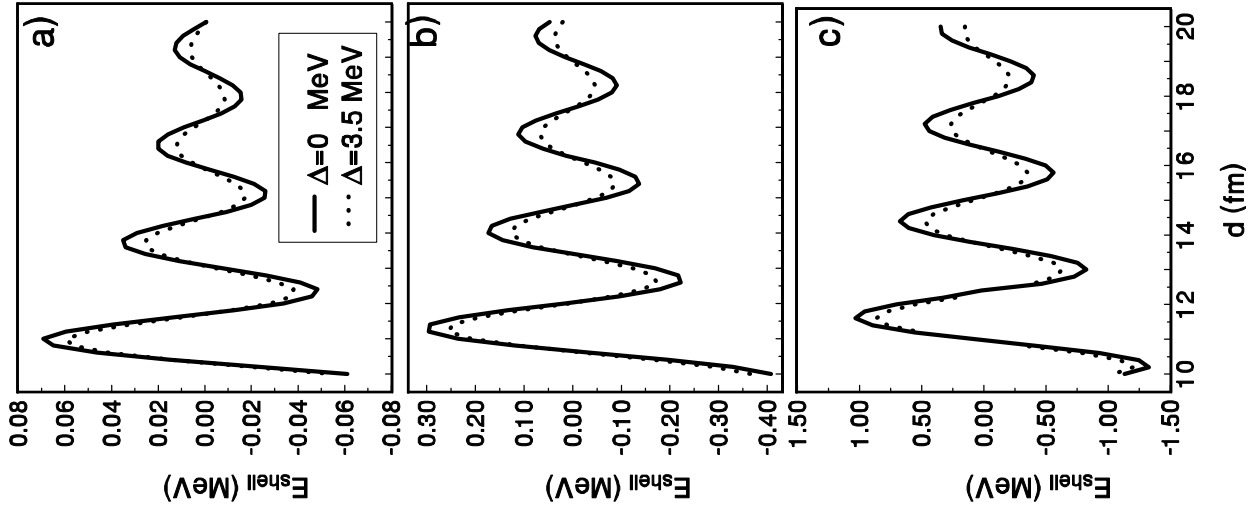
\end{document}